\documentclass[journal=ancac3,manuscript=letter]{achemso}
\usepackage{graphicx}
\usepackage{graphics}
\usepackage{amsmath}
\usepackage{amsfonts}
\usepackage{color}
\usepackage{morefloats}
\usepackage[dvipsnames]{xcolor}

\usepackage{caption}
\usepackage{amssymb}
\usepackage{subcaption}
\usepackage{siunitx}
\usepackage{tikz}
\usepackage[export]{adjustbox}
\usepackage{multicol}

\usepackage{soul}

\newcommand*{\VR}[1]{{\color{OliveGreen}{#1}}}

\def\kB{k_{\mathrm{B}}}
\author{Zhujie Li}
\affiliation[]
{Applied Theoretical Physics-Computational Physics, Physikalisches Institut, Albert-Ludwigs-Universit{\"a}t Freiburg, D-79104 Freiburg, Germany}

\author{Victor G. Ruiz}
\affiliation[]
{Research Group for Simulations of Energy Materials, Helmholtz-Zentrum Berlin, D-14109 Berlin, Germany}
\author{Matej Kandu{\v c}}
\affiliation[]
{Jo{\v z}ef Stefan Institute, SI-1000 Ljubljana, Slovenia}
\author{Joachim Dzubiella}
\email{joachim.dzubiella@physik.uni-freiburg.de}
\affiliation[]
{Applied Theoretical Physics-Computational Physics, Physikalisches Institut, Albert-Ludwigs-Universit{\"a}t Freiburg, D-79104 Freiburg, Germany}
\alsoaffiliation[]
{Research Group for Simulations of Energy Materials, Helmholtz-Zentrum Berlin, D-14109 Berlin, Germany}

\title
{Highly Heterogeneous Polarization and Solvation of Gold Nanoparticles in Aqueous Electrolytes}
\abbreviations{} 
\keywords{gold nanoparticle, polarization, image charge, ion-specific effect, metal--liquid interface, molecular dynamics simulation}

\makeatother

\begin{document}
\begin{tocentry}
\centering \includegraphics[width=1.\textwidth]{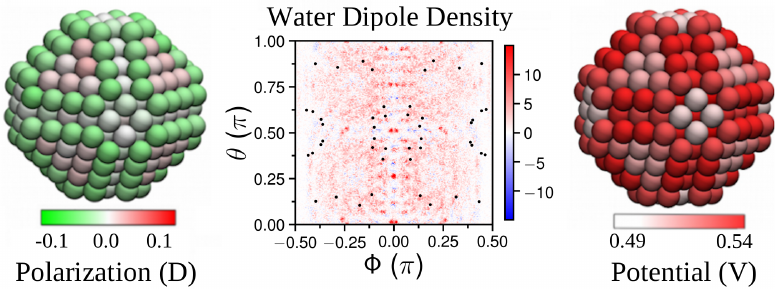}
\end{tocentry}

\begin{abstract}

The performance of gold nanoparticles (NPs) in applications depends critically on the structure of the NP--solvent interface, at which the electrostatic surface polarization is one of the key characteristics that affects hydration, ionic adsorption, and electrochemical reactions. Here, we demonstrate significant effects of explicit metal polarizability on the solvation and electrostatic properties of bare gold NPs in aqueous electrolyte solutions of sodium salts of various anions (Cl$^-$, BF$_4$$^-$, PF$_6$$^-$, Nip$^-$(nitrophenolate), and 3- and 4-valent hexacyanoferrate (HCF)), using classical molecular dynamics simulations with a polarizable core-shell model of the gold atoms. We find considerable spatial heterogeneity of the polarization and electrostatic potentials on the NP surface, mediated by a highly facet-dependent structuring of the interfacial water molecules. Moreover, ion-specific, facet-dependent ion adsorption leads to large alterations of the interfacial polarization. Compared to non-polarizable NPs, polarizability modifies water local dipole densities only slightly, but has substantial effects on the electrostatic surface potentials,  and leads to significant lateral redistributions of ions on the NP surface. Besides, interfacial polarization effects on the individual monovalent ions cancel out in the far field, and effective Debye--H{\"u}ckel surface potentials remain essentially unaffected, as anticipated from continuum `image-charge' concepts.  Hence, the explicit charge response of metal NPs is crucial for the accurate description and interpretation of interfacial electrostatics (as, e.g., for charge transfer and interface polarization in catalysis and electrochemistry).
\end{abstract}

Metal nanoparticles (NPs), in particular gold nanoparticles (AuNPs) have been extensively studied as promising materials for various applications in biomedicine,~\cite{lee2014h,kyriazi2018m,biology,delivery2,cancer2} plasmonic resonance,\cite{plasmon,Nguyen2019} and NP-mediated catalysis.~\cite{enzyme,Wu2012a,catalysis,Roa2017a,mao2020t,ishida2019i,Herves} AuNPs have useful physicochemical properties because of their high surface-area-to-volume ratio as well as distinctive surface morphologies and chemical compositions, which can be very different from those of bulk or cluster states.~\cite{synthesis,defect,Bond2012}  In liquid-phase applications, ionic adsorption at the metal--solvent interface is of utmost importance, not only for the formation of the electric double layer, charge transfer and redox reactions at the metal surface, but also for the synthesis and colloidal stability of metal NP in solutions.~\cite{Merk2014b}  The solvation of AuNPs typically also involves (wanted or unwanted) surface functionalizations with surfactants/ligands.  For a more fundamental understanding and better control of NP properties, it is thus essential to separate ligand and bare AuNP effects on ionic adsorptions. Experimentally, the use of laser-fragmentation or laser-ablation technique has made it possible to synthesize and study ligand-free AuNPs in aqueous solutions,~\cite{Sylvestre,Rehbock,Ziefuss2019,DeAndaVilla2019} among which the anion-specific adsorption on AuNPs is critical for controlling the particle size distribution and colloidal stabilization in solutions.  However, many fundamental processes regarding the interfacial hydration and the adsorption of ions at NPs are still difficult to study on the atomic scale and remain not well understood.~\cite{Wallentine2020d} The structured hydration and ionic adsorption patterns are important for colloidal stability, electrophoretic parameters, and foremost in catalytic and electrochemistry processes as they define details of the channeling charge transfer and can strongly affect electrodeposition and dissolution processes.\cite{Magnussen2019t} 

Classical all-atom molecular dynamics (MD) simulations accelerate our understanding of the microscopic picture of the solid--liquid interface, as they can explicitly and efficiently access the detailed interfacial structure and adsorption mechanisms at surfaces.~\cite{Horinek2007,Horinek2008} 
Importantly, these simulations allow us to explore the effects of the metal polarization in the rearrangement of water molecules by taking fully into account the highly dynamic behavior of the solvent with its significant local variations. The latter is particularly relevant for modeling electrolytes in solution at room temperature. MD simulations have already been exploited to investigate gold--liquid interfaces, such as the studies of halides and biomolecules in water in contact with planar gold,~\cite{Meena2016b,Heinz2016a,Geada2018a} adsorption of water on bare~\cite{Li2015b} or functional AuNPs,~\cite{Heikkila2012,Yang2015a} structure and dispersion state of citrate-coated AuNPs in saline solutions,~\cite{Perfilieva2019a,Franco-Ulloa2020} and peptide adsorption on bare AuNPs.\cite{Shao2018b}  
In particular, by (nonpolarizable) MD simulations we recently revealed substantial ion-specificity and facet selectivity in the adsorption structure and spatial distribution of various ions on AuNP surfaces:
while sodium and some anions (e.g., Cl$^-$, HCF$^{3-}$) adsorb more at the 'edgy' (100) and (110) facets of the NPs, where the water hydration structure is more disordered, other ions (e.g., BF$_4^-$, PF$_6^-$, Nip$^-$(nitrophenolate)) prefer to adsorb strongly on the extended and rather flat (111) facets.~\cite{Li2020a} Moreover, we calculated effective far-field electrostatic surface potentials and for all salts found negative effective surface potentials in the range from $-$10~mV for NaCl down to about $-$80~mV for NaNip, consistent with typical experimental ranges for the zeta potential.~\cite{Li2020a} 

Yet, a shortcoming of most previous studies is the absence of explicit metal polarization and its effects present at the interfaces formed in such systems, e.g., on hydration, adsorption, and charge transfer.~\cite{Tandiana2021} The effects of induced charges in metal surfaces have long been known to be significant in vacuum from an electronic-structure perspective,~\cite{Lang1973} whereas the effects on interfaces with aqueous solutions, biomolecules, and ionic liquids have only started to be addressed recently in a quantitative manner both from the quantum and classical perspectives.~\cite{Heinz2011, Magnussen2019t,Geada2018a}  
But so far, the influence of metal polarization in complex, heterogeneous NP interfaces in ion-specific electrolytes has not been fully scrutinized from an atomistic perspective.~\cite{Tandiana2021}
Noteworthy, recent simulation studies in that direction include a capacitance-polarizability interaction model to study water structure on AuNPs in neat water~\cite{Li2015b} or a rigid-rod model with the GolP-CHARMM force field to study AuNP-citrate interactions.~\cite{Perfilieva2019a} A reactive force field (ReaxFF) MD simulation was employed to study peptide adsorption on AuNPs.~\cite{Monti2019,Samieegohar2019} A computationally efficient polarizable gold model based on a core-shell method was introduced recently by Geada \textit{et al.}, who studied the influence of polarization on the adsorption of charged bio-molecules at the planar gold surface.~\cite{Geada2018a}  However, despite the importance of polarization of gold for interfacial electrostatics,~\cite{Heinz2011,Geada2018a,Ntim2020} the relation between the induced charge on AuNPs and its coupling to hydration and ion-specific adsorption has not yet been systematically addressed. 

From a continuum electrostatic picture, the problem of ionic interaction with metal surfaces has an elegant analytical solution for a metal sphere, based on the concept of ``image charges''~\cite{friedman1975image, bratko1986, Petersen2018d}. On the basis of the continuum model, the attractive interaction of a $z$-valent ion in radial distance $r$ to the metal sphere of radius $a$ is then
\begin{equation}
    w_0(r)/\kB T=-\frac{1}{2}z^2\lambda_\mathrm{B} \frac{a^3}{r^2(r^2-a^2)}
    \label{eq:w0}
\end{equation}
where $\kB$ is the Boltzmann constant, $T$ the temperature, and $\lambda_\mathrm{B}=e_0^2/(4\pi\epsilon\epsilon_0 \kB T)$ is the Bjerrum length, the value of which equals $\lambda_\mathrm{B}\simeq0.72$~nm in water at ambient temperature. At large distances of the ion from the sphere, the image attraction decays rather quickly, as $ r^{-4}$. In the vicinity of the surface, on the other hand,  it is approximately $w_0/\kB T\simeq -z^2\lambda_\mathrm{B}/(4 \Delta r)$, where $\Delta r=r-a$ is the distance from the sphere's surface. From here, it can be easily seen that the image interaction falls below the thermal energy scale ($|w_0|<\kB T$) at distances larger than $\Delta r>z^2\lambda_\mathrm{B}/4$, which for monovalent ions amounts to $\sim 0.2$~nm ---essentially the ionic size. Beyond this distance, in the far field, the explicit effect of polarizability should be negligible, at least for low-valency ions. 

Thus, from both the microscopic and continuum pictures, the aspect of polarizability should be included in classical modeling for studying surface-related phenomena such as ionic adsorption. Furthermore, it also becomes clear that the polyhedral geometry of metal nanoparticles and their atomistically structured surfaces become of utmost significance when considering polarization, which must consequently be quite heterogeneous.  The aim of this work is thus to systematically investigate the effects of explicit polarizability on hydration and ionic adsorption, and how these phenomena are coupled. As in our previous reference work using non-polarizable MD simulations,~\cite{Li2020a} we consider ions of various shapes and valencies that are relevant to potential applications and compare the results to those without the image charge effect.  
Our studied anions and AuNPs are exemplified in Figure~1.  We employ the polarizable core-shell model introduced by Geada {\it et al.}~\cite{Geada2018a} This polarizable model is simple enough, fully compatible with common biomolecular and materials-oriented force fields, and more importantly, it was developed following the interface force field (IFF) approach, ~\cite{Heinz2013b,Geada2018a} which was used in our previous study on nonpolarizable NPs.~\cite{Li2020a}

\begin{figure}[ht]
	\centering
	\includegraphics[width=1\linewidth]{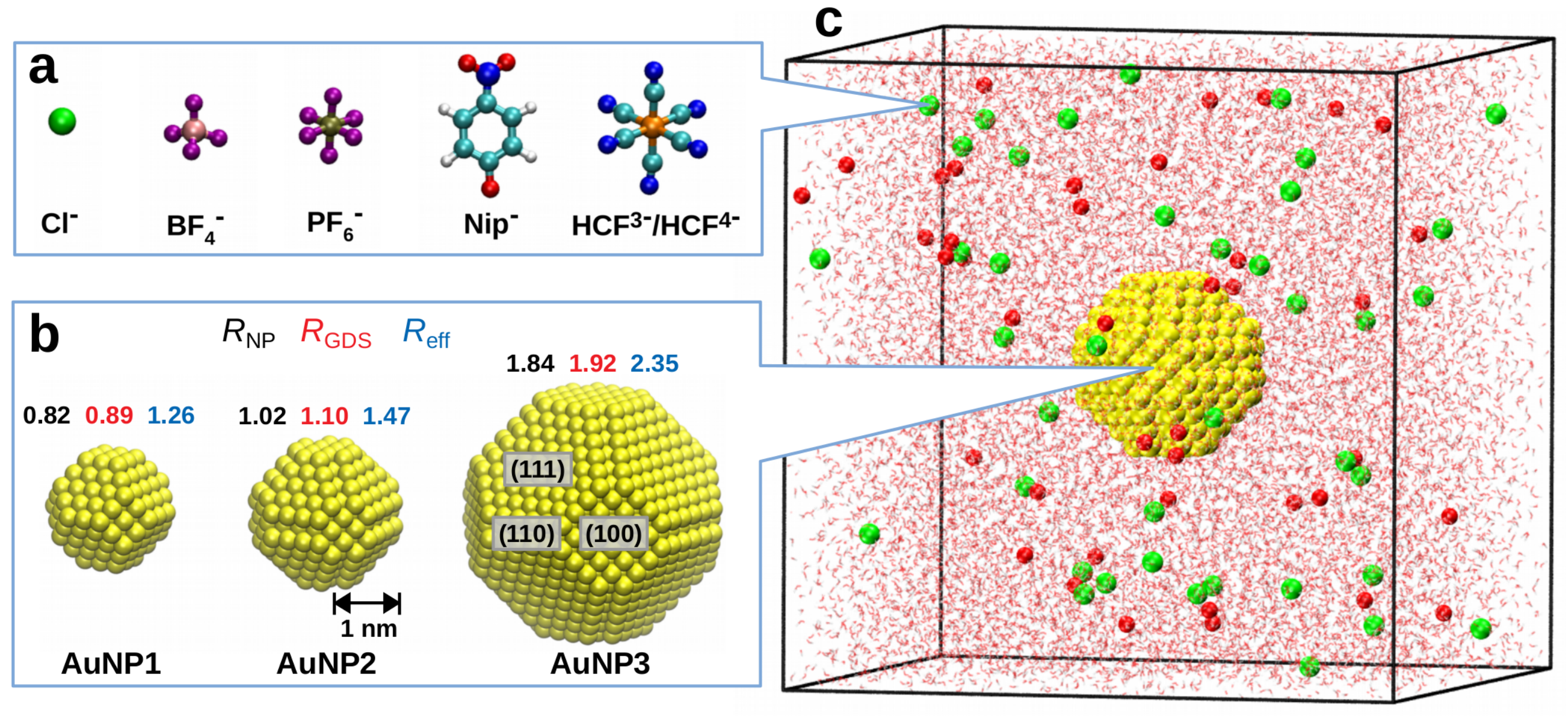}
	\caption{Models and simulations. (a) Anions studied in this work:
	chloride (Cl$^{-}$), tetrafluoroborate (BF$_4^-$), hexafluorophosphate (PF$_6^-$), nitrophenolate (Nip$^-$), ferricyanide (HCF$^{3-}$), and ferrocyanide (HCF$^{4-}$). (b) Simulated gold nanoparticles (AuNPs) of three different sizes, where $R_{\rm NP}$ stands for the bare radius (defined as the distance to the farthest facet center ({\it i.e.}, (100) facet center) from the center of the NP),  $R_{\rm GDS}$ represents the radius of the Gibbs dividing surface, and $R_{\rm eff}$ defines the radius of the effective Debye--H{\"u}ckel surface (see text), all in the unit of nm. (c) Simulation cell with AuNP placed in the center and surrounded by Na$^+$ cations (red), representative Cl$^-$ anions (green), and water molecules (shown in the background).}
	\label{fig:box}
\end{figure} 
 
\section{RESULTS AND DISCUSSION}

\begin{figure}[h!]
	\centering
	\includegraphics[width=1.\linewidth]{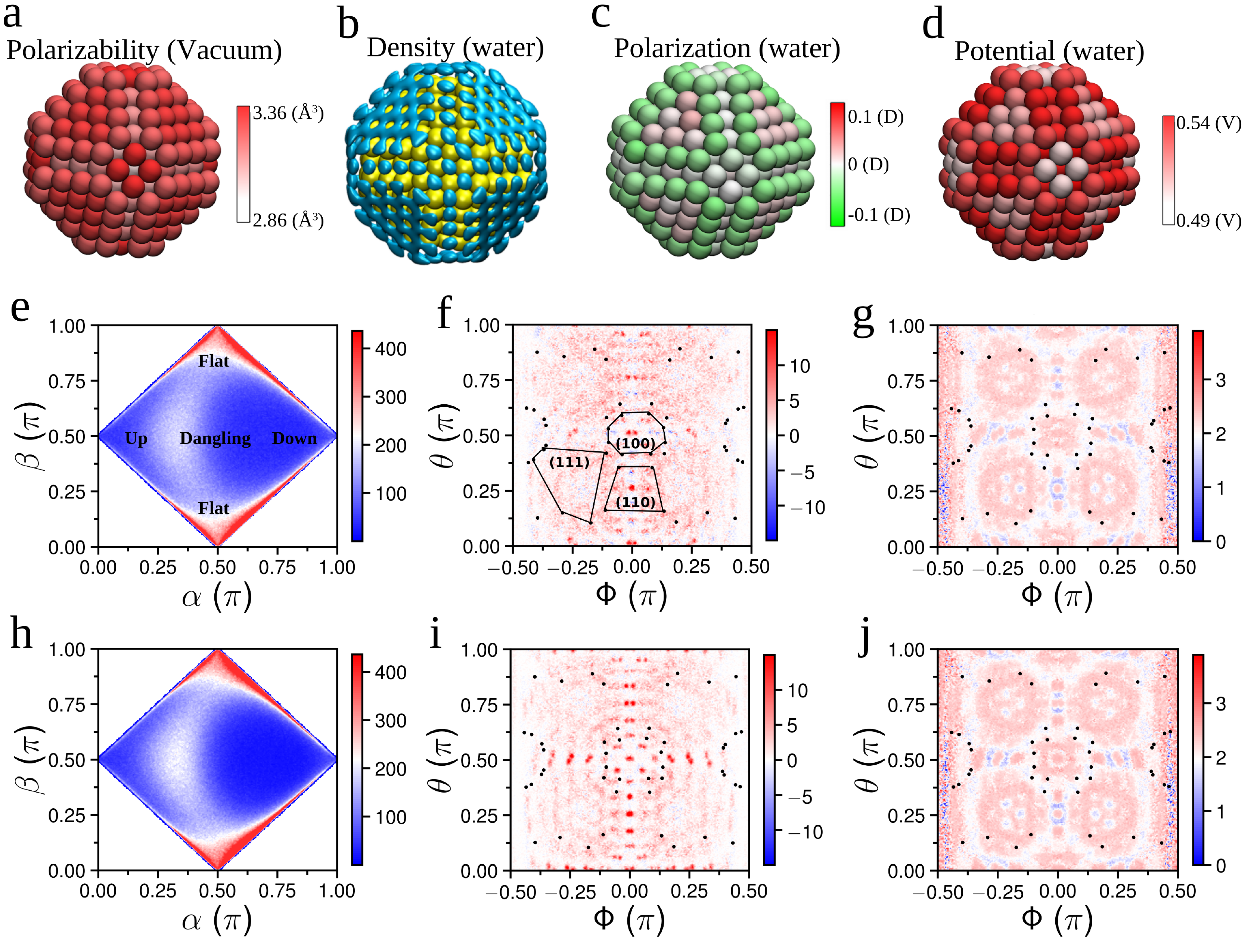}
	\caption{(a) Intrinsic polarizability of Au atoms for AuNP2 in vacuum. (b)--(d) Properties of AuNP2 in pure water: (b) Spatial density distribution of water (shown by an iso-density plot of oxygen atoms), (c) radial dipole moment of the Au atoms in the unit of Debye (D), (d) electrostatic potential at each Au atom's center. See text and Methods for definitions of polarization and potential. (e)--(g) Hydration properties at polarized AuNP2: (e) Water orientation at the NP interface, shown by the intensity of a 2-dimensional histogram of the two water angles (see text for the definition of these angles), (f) contour plot of the angle-resolved water radial dipole density ($\Phi$ and $\theta$ denote the azimuthal and polar angle, respectively, in the spherical coordinate system), and (g) average number of water--water H-bonds per water molecule. 
	(Black dots in (f), (g), (i), and (j) are the positions of vertex atoms in the NP. Typical facet regions are depicted in (f) for orientation). (h)--(j) Corresponding hydration properties at nonpolarizable AuNP2. Only the interfacial water molecules (within the first solvation shell of the AuNP2) were taken into account for panels (e)--(j).}
	\label{fig:polarizeWater}
\end{figure}

\textbf{AuNPs in Neat Water}. We first take a look at polarizable hydration properties of a AuNP in neat water.  Details of the simulations and analysis methods can be found in the Methods section. As an interesting reference, Figure~\ref{fig:polarizeWater}a shows the intrinsic polarizability (see Equation~\ref{eq:mu} in Methods) of the AuNP2 atoms in vacuum. We observe heterogeneity in the polarizability of the surface atoms, for example, the (100) facet is most polarizable. This is expected because dipole fluctuations should be enhanced in atoms with fewer neighbors. Hence, because of this heterogeneous intrinsic polarizability, also the polarization response is expected to be heterogeneous.  When the NP is solvated by water (see the iso-density plot in Figure~2b), water molecules induce a distinct polarization pattern on the nanoparticle's surface. In Figure \ref{fig:polarizeWater}c the gold atoms are colored in terms of their respective radial dipole moments (dipole vector per atom projected on the radial vector from the center of the NP, see Equation~\ref{eq:mu} in Methods). Positive values (red shades) depict dipoles pointing outward and negative values (green shades)  pointing inward to the NP.  Specifically, the `corner' (100) facets and the 
`edgy' (110) facets exhibit mostly negative dipole moments, whereas the extended (111) facets are positively polarized. This polarization pattern is induced by the hydration structure, as indicated by the iso-density plot in Figure~2b and discussed in more detail below. Figure~\ref{fig:polarizeWater}d shows the electrostatic potential at surface Au atoms (see Equation~\ref{eq:pot} in Methods).
The lowest potential ($\approx$~0.49 V) is observed on the (100) facets, higher on the (111) facets, and the highest ($\approx$~0.54 V) on the (110) facets.  Hence, the overall relative modulation of the surface potential due to the heterogeneous hydration (and polarization) is within 10~$\%$. Electrostatic potential heterogeneity in the surrounding of polarizable gold nanoparticles in pure water has also been reported in a recent simulation work.~\cite{Perfilieva2019a}

The reason for this heterogeneous polarization stems from the highly facet-specific hydration structure of interfacial water. Gold atoms at the (100) and (110) facets have fewer neighbors than the (111) facet. Such a distinct coordination environment is associated with a heterogeneous hydration structure of AuNP, as can be seen in Figure~\ref{fig:polarizeWater}b, which features local density depletions at the edges and corners, indicating defective water--water hydrogen-bond structures or other mismatches in the structure of surface water. Water molecules are expected to structure differently according to the underlying pattern of the specific surface of the NP. This is also supported by previous results on planar gold surfaces~\cite{Geada2018a} as well as electronic-structure calculations of a single water molecule on Au(111) and Au(100) surfaces.~\cite{LinGross2012}  

In order to inspect the microscopic structure of interfacial water at polarized AuNPs in more detail, we analyzed the water orientation with respect to the NP (Figure~\ref{fig:polarizeWater}e), which can be determined by two angles, $\alpha$ and $\beta$, as proposed in a recent work~\cite{Tandiana2021}. Angle $\alpha$ is defined as the angle between the dipole of the water molecule and the vector formed between the center of the NP and the oxygen atom of the water molecule; whereas $\beta$ denotes the angle between the normal of the water molecule and the vector formed between the center of the NP and the oxygen atom of the water molecule; see the schematic diagram in Figure S1 in the Supporting Information. According to this definition, the criteria for the `up' and `down' configurations are $\alpha < 0.4\pi $ and $\alpha > 0.6\pi$, respectively, regardless of $\beta$\VR{;} the flat configuration is characterized by $0.45\pi < \alpha < \frac{2}{3}\pi$ and either $\beta < \frac{1}{6}\pi$ or $\beta > \frac{5}{6}\pi$; and the dangling configuration corresponds to the remaining configurations. As can be seen in Figure~\ref{fig:polarizeWater}e, a large proportion of water molecules lie flat on the NP, a somewhat smaller fraction adopts a configuration that is in between the up and dangling configurations, while only a small fraction adopts the down configuration.  

Further analysis of the local polarization of the interfacial water also confirms these observations (Figure~\ref{fig:polarizeWater}f), where we plot the water dipole moment density in radial direction from the NP center as a function of $\Phi$ and $\theta$, the azimuthal and polar angle, respectively, in the spherical coordinate system of the NP. The dipoles exhibit values close to zero (corresponding to flat configurations in Figure~2e) or slightly positive values (the up and dangling configurations). The angle-resolved contour plot in Figure 2f also demonstrates again the facet selectivity of water structure in terms of radial orientation, where the radial dipole moments of water have larger values on the (100) and (110) facets than on the (111) facets. In addition, we present the  average number of water--water hydrogen bonds (H-bonds) (defined by a donor--acceptor (D--A) distance of less than 3.5 \AA~ and a D--H--A angle larger than 150 degrees~\cite{luzar1996hydrogen}) per water molecule in Figure 2g. While the average number of H-bonds per water molecule in bulk is 2.6, it decreases down to 2.1 for water molecules at the interface. More precisely, the H-bond formation is facet-dependent, with the values around 2.2($\pm$0.3) on the (100) and (111) facets while lower values around 1.8($\pm$0.3) appear on the (110) facets.  Hence, a distinct facet-specific hydration structure couples to the heterogeneous polarization and potential distributions shown in plots Figures 2c and 2d. 

By comparing the above outcomes to our nonpolarizable simulations,\cite{Li2020a} we find that the explicit polarizability has small effect on the hydration structure: Comparisons of orientations, dipoles, and H-bond structures are shown in Figures 2h, 2i, and 2j, respectively. If anything, small changes in the local dipole structures are observed (Figure~2i) pointing to subtle, local reorientations of water molecules at certain surface sites. This must be viewed as a coupled response to the polarization and potential modulation of the AuNP, cf. Figure 2c and 2d, as induced by the hydration structure itself. In other words, hydration structure is dominated by the NP geometry and surface, and polarizability leads to perturbation of the hydration structure within a self-consistent response of the interfacial electrostatics, which is in agreement with the observation that the hydrogen bond network is insensitive to the polarizability of similarly polarizable surfaces.~\cite{ho2013polar} We will see later that these small changes in water dipoles induce significant modifications of the electrostatic potential distribution also in the radial direction from the NP. 

\begin{figure}[h!]
	\centering
	\includegraphics[width=1.\textwidth]{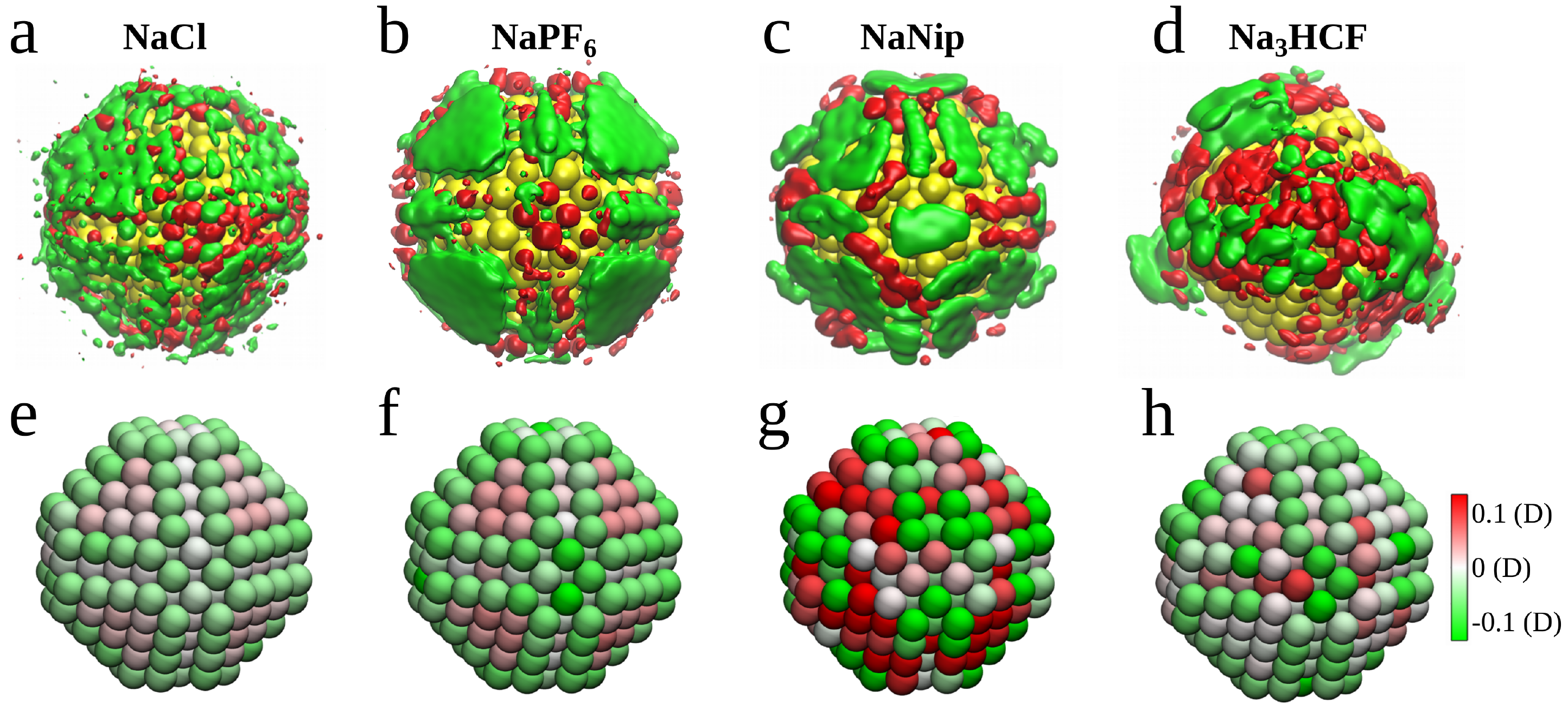}
	\caption{Ionic adsorption and polarization of AuNP2. (a)--(d) Spatial density distributions of ions (cations in red, anions in green) around the AuNP2 shown by iso-density surfaces for representative systems. Iso-density values are set to 0.5, 0.8, 6.0, 1.0 nm$^{-3}$ in (a)--(d), respectively. (e)--(h) Polarization (radial dipole moment) of the NP surface atoms in the presence of adsorbed ions shown above.}
	\label{fig:sdf-dipole}
\end{figure}

\textbf{Ion-specific Adsorption and Its Effects on Polarization of AuNP}. We now turn to ionic adsorption in different electrolytes and their influence on the surrounding water structure. 
In Figure~\ref{fig:sdf-dipole}a--d we show spatial distribution of ions in the form of iso-density surfaces for a few representative systems. Cations are presented in red and anions in green. As seen, the figures indicate clear facet selectivity for ion adsorption. Specifically, Na$^+$ cations preferentially bind to the ``edgy'' (100) and (110) facets. Cl$^-$ anions, on the other hand, prefer the adsorption on the (111) facets, forming a dense Cl$^-$ monolayer. The situation is similar for NaPF$_6$ but with a more distinct adsorption pattern between cations and anions, owing to the higher adsorption.\cite{Li2020a} The behavior is different for Nip$^-$ and HCF$^{3-}$ where irregular, not very symmetric adsorption patterns appear. These ions adsorb very strongly and the ion distributions do not seem to be sampled well within our simulation times. In other words, local residence times of anions adsorbed at specific sites are comparable or longer than the simulation time of about tens to hundreds of nanoseconds. 
The adsorption of HCF$^{3-}$ anions is substantially enhanced by polarizability (compare to the non-polarizable model in Figure S2 in the Supporting Information), which can be explained by the square dependence of the image-charge attraction with the valency, $w_0\propto z^2$ (see Equation \ref{eq:w0}).

In turn, the adsorption of ions can affect the polarization of gold in an ion-specific way, as shown in the bottom row of Figure~\ref{fig:sdf-dipole}.  On the one hand, the adsorption of NaCl ions (Figure~\ref{fig:sdf-dipole}e) cause only moderate changes when compared to neat water (Figure 2c), with a visible difference mostly at the (100) facet, where the cations like to adsorb. 
Larger changes are found for the more strongly adsorbing NaPF$_6$ ions, which tend to polarize the surface more, while not changing the qualitative picture (i.e., the overall polarization pattern is the same as in neat water). Specifically, negative dipole moments gather mostly at the `corners' and `edges' of the NP, while positive ones are observed on the (111) facets. 
On the other hand, the situation changes completely in the NaNip and Na$_3$HCF systems because of the irregular and strong adsorption of these ions, leading to locally very high polarizations, very different to the neat water case. (A more detailed, facet-resolved analysis can be found in Figure S3 in the Supporting Information).

Our analysis demonstrates that complex adsorption patterns and facet selectivity of adsorbing anions lead to a very heterogeneous polarization of the AuNP. Namely, ions and charged molecules have preferential adsorption sites, which are coupled to various interaction effects. A dominant one, as discussed above, is the heterogeneous hydration structure (\textit{i.e.,}, the role of interfacial water~\cite{Argyris2010}), which is already present in nonpolarizable systems\cite{Li2020a}. The induced polarization of the NP surface due to the presence of the solution, cf. Figure~\ref{fig:polarizeWater}c and Figures 3e-h, in turn, modifies the local water and ionic structure. As can be seen from Table~\ref{tbl:facet-cn}, for example, the averaged adsorption numbers of Cl$^-$ anions per gold atom at all surface sites are larger than those of Na$^+$ cations with and without image charge on the AuNP2. Importantly, the inclusion of the image-charge effect significantly changes the adsorption ratio between anions and cations at corner (100) and edgy (110) facets, with an increasing proportion of adsorbed cations. On the other hand, the image-charge effect has a minor impact on the (111) facets. Hence, polarization has a significant effect on the interfacial {\it lateral} ionic distributions on the surface.

\begin{table}[h!]
    \centering
    \caption{Ratio between the adsorbed anions and cations on gold atoms on each of the facets for the AuNP2-NaCl system (See Methods for the calculation of the adsorption numbers).}
    \label{tbl:facet-cn}
    \begin{tabular}{ccccccc}
        \hline 
		\rule{0pt}{3ex} &(100)&(110)&(111)  \\
		\rule{0pt}{3ex} Polarizable & 2.9 & 4.0 & 4.2 \\
		\rule{0pt}{3ex} Nonpolarizable & 6.0 & 7.6 & 5.0 \\
		\hline
    \end{tabular}
\end{table}

\begin{figure}[h!]
	\centering
	\includegraphics[width=1.\linewidth]{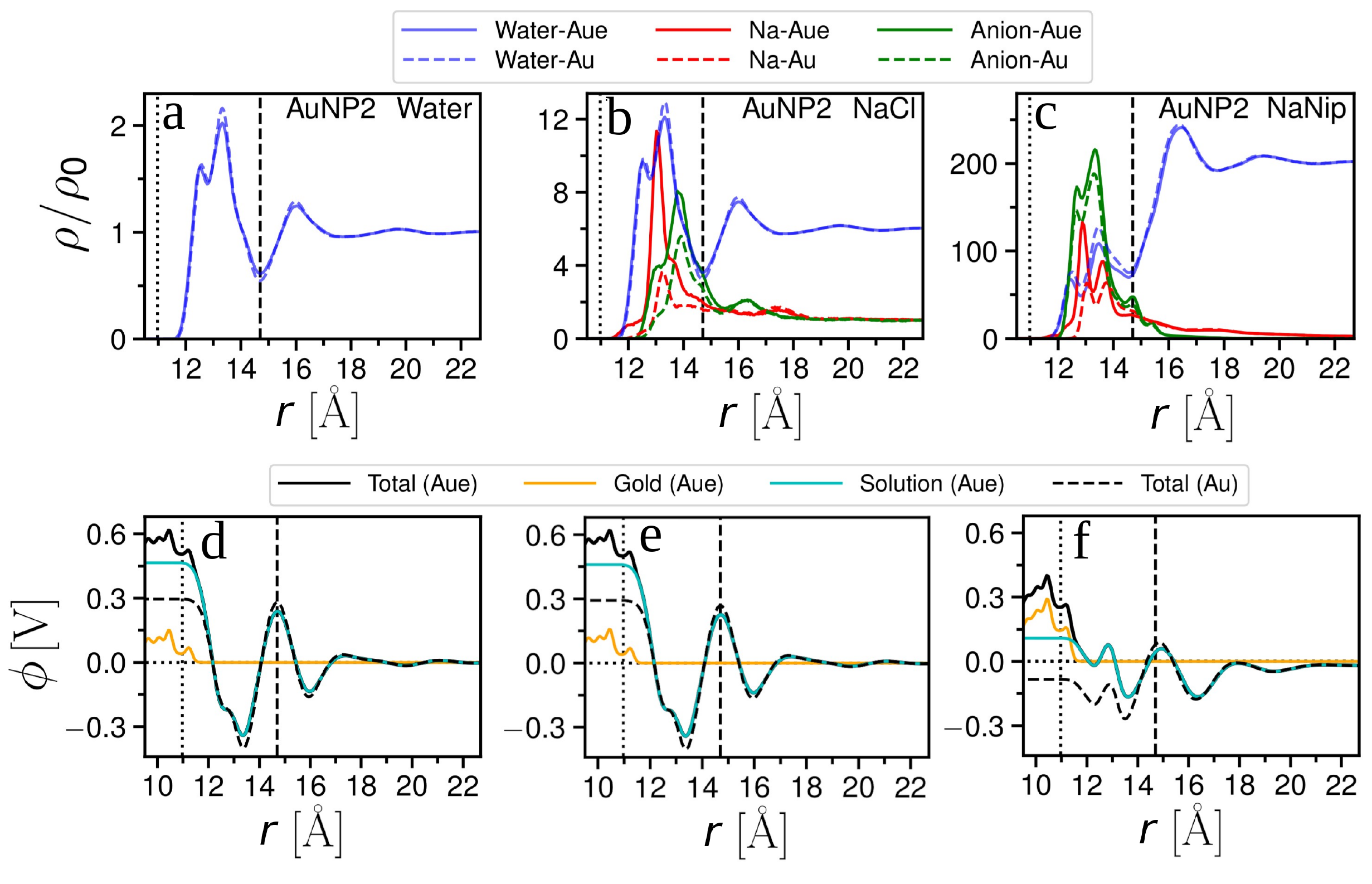}
	\caption{Radial number density normalized by its bulk value as a function of the radial distance $r$ from the NP center for AuNP2 (a) in neat water, (b) in NaCl, and (c) in NaNip electrolytes. Water profiles are scaled by a factor of 6 and 200 in (b) and (c), respectively, for clarity. (d)--(f) Total and decomposed electrostatic potential profiles for the corresponding systems in (a)--(c). Solid and dashed lines represent the results obtained from the polarizable (labeled as `Aue') and nonpolarizable (labeled as `Au') gold model, respectively. The radii of the Gibbs dividing surface ($R_{\rm GDS}$) and effective Debye--H\"uckel surface ($R_{\rm eff}$), calculated as in Ref.~\citenum{Li2020a}, are shown by vertical dotted and dashed lines, respectively. }
	\label{fig:rdf-pot}
\end{figure}

\textbf{Radial Density and Electrostatic Potential Distributions}. Figure~\ref{fig:rdf-pot}a shows normalized density profiles (i.e., angular averaged around the NP) of water in the polarizable AuNP2 (solid lines, `Aue' labels) as a function of the radial distance from the NP center. For comparison, we also plot the values from the nonpolarizable model (`Au' labels) as dashed lines. The water density profile is quite similar in both models. This is not unexpected because both models are parameterized such that they reproduce correct hydration properties of gold surfaces (i.e., Au--water interface tension, hydration energy).~\cite{Geada2018a} However, we observe subtle dipolar rearrangements upon polarization (cf.\ Figure~2f and 2i) and we will show later on that those are sufficient to induce significant changes in the radial electrostatic potential. 

When it comes to ions, the effects of explicit polarizability becomes more significant in the density profiles. As seen from Figure~\ref{fig:rdf-pot}b (see also Figure S4 in the Supporting Information), both Na$^+$ and Cl$^-$ ions are attracted to the AuNP, but the attraction is evidently stronger in the polarizable model. This we attribute to the attractive image interaction in the polarizable model, which is lacking in the nonpolarizable one. For instance, the adsorption peak of Na$^+$ is by a factor of 3 higher in the polarizable model. The difference is smaller for Cl$^-$, but its density profile is broader. Since the Na$^+$ cation is a bit smaller than the Cl$^-$ anion, the former approaches the gold surfaces more, as clearly seen from Figure~\ref{fig:rdf-pot}b.  In fact, Na$^+$ cations approach the surface as close as water molecules, implying that they get (partially) dehydrated and directly contact the gold atoms, which is in accord with observations at planar gold surfaces~\cite{Geada2018a} and surfaces with localized charge.~\cite{Dewan2014}.

As we exchange Cl$^-$ ions by Nip$^-$ in the electrolyte (Figure~\ref{fig:rdf-pot}c), we witness several important changes. Since Nip$^-$ has a large surface area that epitaxially fits the gold surface (in particular the phenyl group), it tends to successfully form many van der Waals contacts with the surface. Therefore, the Nip$^-$ ion adsorbs considerably more strongly to a AuNP than the Cl$^-$ ion. Since van der Waals forces are not sensitive to image charges, the difference between both polarizable and nonpolarizable models is smaller than for monatomic ions. Moreover, many adsorbing Nip$^-$ ions to the nanoparticle drag additional Na$^+$ counterions to the surface in order to partially compensate the electric charge. This synergetic adsorption, however, is not perfectly electroneutralizing, and a negative charge dominates within the first hydration layer. 
Density profiles and charge distributions for other studied ions are provided in the Supporting Information (Figures S5 and S6--S8, respectively) and exhibit a similar trend as the above two representative salt systems.

Figure~\ref{fig:rdf-pot}d--f shows the total radial electrostatic potential around AuNP2, decomposed into the contributions stemming from individual species (gold: Au core and shell; solution: water O and H, and/or ions) in pure water, NaCl, and NaNip.  We compare again the polarizable model (`Aue') to the nonpolarizable one (`Au'). The differences between the two models are clearly visible mostly in the region smaller than the effective Debye--H\"uckel radius, $R_{\rm eff}$.  In pure water, for instance, the total potential in the polarizable model around the Gibbs dividing surface, $R_{\rm GDS}$, is by about 0.15--0.2~V higher than in the nonpolarizable one. Around 0.05--0.1~V of this difference comes from the gold atoms in the polarizable NP (see Figure~\ref{fig:rdf-pot}d). 
Hence, although interfacial water structure overall seems to be quite indifferent to the addition of explicit polarization, small dipole rearrangements due to polarization have a significant effect on the interfacial potential. Adding NaCl to the solution (Figure~\ref{fig:rdf-pot}e), does hardly change the total potential because the NaCl adsorption is low. The situation is different for the strongly adsorbing NaNip, Figure~\ref{fig:rdf-pot}f, where all contributions change considerably. Note that the total potential at $R_{\rm GDS}$ even changes the sign when going from the nonpolarizable situation (negative potential in gold) to the polarizable one (positive, as in the neat water and NaCl cases). Because of the lack of theoretical and experimental evidence, it is difficult to assess how quantitative these predictions are, but they clearly highlight the importance of metal polarizability for strongly adsorbing ions and their possible effect on local (redox) potentials. 

Furthermore, by separating water and ionic contributions to the electrostatic potential (Figure S9 in the Supporting Information), we find that the ionic contribution to the total potential is higher in the polarizable model by $\approx$ 0.1~V, which can be  associated to the presence of more cations at polarizable gold surfaces, whereas the water contribution decreases.  Hence, water and ionic contributions are balancing each other, while polarizability influences the balance.   In fact, classical non-polarizable MD studies have found that the rearrangement of water molecules at the interface is actually disrupted by the presence of ions,~\cite{Tandiana2021} showing that the proportion of dangling water molecules change according to the size of the NP and the concentration of ions. This has consequences for water hydrolysis in the presence of gold NPs as an example.  Experiments have demonstrated that adding 0.1\% (m/v) of NaCl into solution slows down the production of hydroxyl radicals, which can be explained by the rearrangement of water molecules due to ions.~\cite{Gilles2018} 

Importantly, the total potential difference between nonpolarizable and polarizable models becomes negligible at large distances for both electrolytes (see also Figures S10 and S11 in the Supporting Information for other systems). This indicates that image charges indeed play an important role only within the first water layer and for individual ions,  but have little influence for the far field for the monovalent systems, as already implied by the continuum model.~\cite{Petersen2018d} This has consequences on the effective surface charges and potentials obtained from fitting Debye--H{\"u}ckel theory to the potential distributions in the far field, as reported previously for the nonpolarizable case.~\cite{Li2020a} Comparing to the polarizable model (Table S1 in the Supporting Information), the overall effective charges and potentials are negative and have similar results within statistical errors for both AuNP models with and without image charges. This agrees with the above observation of no obvious potential difference beyond the effective Debye--H{\"u}ckel radius, i.e. ionic interface effects due to polarizability cancel out in the far field. (The effective charge values are defined with respect to an effective Debye--H{\"u}ckel radius, $R_{\rm eff}$, which we set as the location of the first minimum of the water density distribution, see Figure 4 and our previous works.\cite{Li2020a,Xu2017c,Nikam2018}) However, the tri- and tetra-valent systems (AuNP2-Na$_3$HCF and AuNP2-Na$_4$HCF, respectively) reveal a sizeable deviation in effective surface charges between the two gold models. Here, effective charges and potentials are larger in absolute values in the polarizable model. This is reasonable because the multivalency enhances ionic adsorption in the presence of image charges, and polarizability induced changes are amplified and propagate visibly into the far field. In summary, the image-charge effects are important for near-field properties right at the interface (e.g., adsorption structure, surface potential), while becoming much less significant for far-field properties for at least all studied monovalent salt systems in this work.

\section{CONCLUSION}

We have performed MD simulations with a polarizable gold model to study hydration and ion-specific adsorption in presence of polarizability (``image-charge" effects) and its consequence on the polarization of AuNPs. We compare the solvation and double-layer structure of three different bare AuNPs of radius between 1--2 nm with and without gold polarizability in sodium-based aqueous electrolytes with various anions. Image-charge effects on ionic adsorption at solid--liquid interface were highlighted by enhanced density distribution together with the heterogeneity, among which the adsorption of Na$^+$ cations revealed that they are more attracted to the surface with image charges and consequently leaving some direct contacts with gold atoms on the edges and corners of AuNPs. The resulting electrostatic potentials exhibit significant dependence on the image charges in the near field (at the interfacial region), however, there is no obvious influence on the far field according to the calculated effective potentials. Herein, one important conclusion can be drawn: the inclusion of image-charge effects in AuNP is essential for near-field properties right at the interface since these play a key role in electrochemical and catalytic processes, while it does not change the far-field properties, such as the colloidal stability, for all studied monovalent salts in present work. 

Importantly, the polarizable model gives access to study the polarization of AuNPs in different complex environments, including molecular ions and water, which is crucial for charge transfer at the NP--liquid interface in electrochemical reactions. Significant ion-specific effects on the induced polarization were observed along with heterogeneous distributions, which can be traced back to the adsorption of ions in terms of charge densities and binding sites. This work sheds light on many fundamental aspects of the microscopic picture of NPs in aqueous electrolytes, which deepens the understanding of these systems, complement experiments, and could be useful for future rational design of functional NPs.  

We encourage future studies on the development of better polarizable force fields, including the polarizability of water.~\cite{Rozsa2021} One could benchmark classical approaches to {\it ab-initio} MD (AIMD) approaches and compare water oxygen and hydrogen profiles as well as the total charge contribution across the interface.~\cite{SakongGross2018,Le2017,heenen2020solvation} AIMD studies suggest, for instance, that the interface dipole has significant contributions from partial charge-transfer from the water layer to the metal, given that there is no preferential orientation of the solvating water molecules in metal--water interfaces.~\cite{SakongGross2018}. 
Effects of charge-transfer at the gold--water interface have been found to be smaller in comparison to other metal surfaces~\cite{Clabaut2020b}, but a full quantitative understanding (including the presence of ions) is still not achieved. A consistent convergence of AIMD and classical polarizable models in that matter is highly desirable. Intimately related to this is  the quantification of the work function, which is tuned by dipolar contributions present at the interface,~\cite{SakongGross2018} and, therefore, an important quantity to study and control for applications. 

\section{METHODS}
\textbf{Models}. We employ all-atom MD simulations to investigate AuNP in aqueous solutions. The simulation cells, as shown in Figure~\ref{fig:box}, were built based on the protocol  described in a previous study~\cite{Li2020a}. We used the same force field parameters for water molecules and ions as in our previous work,~\cite{Li2020a} where SPC/E water model was employed~\cite{Berendsen1987} and ions were taken from Refs.~\citenum{Geada2018a,CanongiaLopes2006a,Kanduc2017c,Prampolini2014a}. For the polarizable gold model we used the recently proposed force field by Geada \textit{et al.}~\cite{Geada2018a}. The results were compared to the results obtained from a nonpolarizable model published recently.~\cite{Li2020a} Both the polarizable and nonpolarizable force field for gold were developed by following the interface force field (IFF) approach~\cite{Heinz2013b} and thus validated by various interfacial properties, such as the surface energy, hydration energy, and surface tension with water~\cite{Heinz2008b, Geada2018a}. The improvement of the polarizable model was achieved by explicitly introducing the image charge through a core-shell like model,~\cite{dick1958t,lemkul2016e} which is crucial to the adsorption of charged species onto gold surfaces.

\textbf{Simulations}. The simulations were performed using the LAMMPS simulation package~\cite{Plimpton1995} with an integration time of 1 fs. The temperature in the simulations was $T$ = 298.15 K and maintained by Nos{\'e}--Hoover thermostat~\cite{hoover1985c} with the time constant of 1 ps. Each simulation was first equilibrated for several nanoseconds in the NPT ensemble using the Parrinello--Rahman barostat~\cite{parrinello1981p} at a pressure of $P$ = 1 bar and with the time constant of 1 ps, followed by 200--300 ns of production runs in the NVT ensemble. Gold atoms were allowed to move during the simulations and the mechanical stability of the AuNPs structure was confirmed in all systems. Simulation details of the studied systems are given in Table~\ref{tbl:cell}. Spatial distribution functions were calculated by TRAVIS program~\cite{Brehm2011a} and visualization of all snapshots were made by VMD software.~\cite{Humphrey1996}   

\begin{table}
	\centering
	\caption{Details of simulated systems for three studied nanoparticles AuNP1, AuNP2, and AuNP3 with various salts. $N_{\alpha}$ ($\alpha$ = cation or anion) is the number of species $\alpha$, and $I$ (= $0.5 \sum_{i} \rho_i^0 z_i^{2}$) is bulk ionic strength (Aue: polarizable gold, Au: non-polarizable gold). $^a$Results taken from our previous work.~\cite{Li2020a}} 
	\label{tbl:cell} %
	\begin{tabular}{ccccccccccc}
		\hline 
		\rule{0pt}{3ex}NPs&salt&$N_{\rm{cation}}$&$N_{\rm anion}$&$L_{\rm cell}$  [nm]&$I$ [$\rm{mM}$] &$I$ [$\rm{mM}$]\\
		\rule{0pt}{1ex} &&&&&(Aue)&(Au)\\
		\hline %
		\rule{0pt}{3ex} AuNP1&NaCl     &22 &22 &6.54&124&130$^a$ \\
		\rule{0pt}{3ex} AuNP2&NaCl     &43 &43 &8.17&127&131$^a$  \\
		\rule{0pt}{3ex} AuNP3&NaCl     &70 &70 &9.68&123&129$^a$ \\
		\rule{0pt}{4ex} AuNP1&NaNip    &35 &35 &6.58&89 &92  \\
		\rule{0pt}{3ex} AuNP2&NaNip    &60 &60 &8.21&85 &93$^a$ \\
		\rule{0pt}{3ex} AuNP3&NaNip    &120&120&9.74&76 &97  \\
		\rule{0pt}{4ex} AuNP2&NaBF$_4$ &43 &43 &8.18&113&119$^a$  \\
		\rule{0pt}{3ex} AuNP2&NaPF$_6$ &43 &43 &8.19&103&113$^a$  \\
		\rule{0pt}{3ex} AuNP2&Na$_3$HCF&45 &15 &8.18&201&249$^a$  \\
		\rule{0pt}{3ex} AuNP2&Na$_4$HCF&40 &10 &8.17&239&287$^a$  \\
		[1ex]\hline
	\end{tabular}
\end{table}

\textbf{Dipole Moment and Polarizability}. The dipole moment of individual gold atoms, which measures the separation of positive and negative charges, was calculated and projected on the radial distance as following
\begin{subequations}
	\begin{align}
	\vec{\mu} &= q \cdot \vec{d}_{\rm{Au-e}} \\
	\langle \mu_r \rangle &= \Big \langle \frac{\vec{\mu}\cdot\vec{r}_{\rm{Au}}}{|\vec{r}_{\rm{Au}}|} \Big \rangle \\
	\alpha_V &= \frac{\langle \mu_r^2 \rangle}{4\pi\epsilon_0k_BT}
	\end{align}
	\label{eq:mu}
\end{subequations}
where $q$ and $\vec{d}_{\rm{Au-e}}$ are the charge and distance vector of the dipole, $\vec{r}_{\rm{Au}}$ is the vector of Au core with respect to the NP center, $\vec{\mu}$ and $\mu_r$ are the dipole moment vector and its radial projected value, respectively, $\alpha_V$ denotes the polarizability in common volume units, and the angle brackets $\langle ... \rangle$ represent the time average. The polarizability is proportional to the mean of the square of dipole moment (\textit{i.e.}, $\langle \mu_r^2 \rangle$ in Equation~\ref{eq:mu}) in the framework of fluctuation--dissipation theorem.~\cite{fornes1998fluctuation} We plot it in Figure 2a in units of \AA$^3$, the calculated values ranged from 2.86 to 3.36 for individual gold atoms in the AuNP2. As a reference, the polarizable model implemented in this work gives a value of 3.33 \AA$^3$ for an isolated gold atom.~\cite{Geada2018a}

\textbf{Electrostatic Potential on Gold}. The electrostatic potential at each of the gold atoms' center is given by,
\begin{equation}
    \phi_i = \frac{1}{4\pi\epsilon_0}\sum_{j, j\neq i}\frac{q_j}{r_{ij}}
    \label{eq:pot}
\end{equation}
where $\phi_i$ is the electrostatic potential at the $i$th gold atom's center, $q_j$ is the charge of the $j$th atomic species, and $r_{ij}$ represents the distance between the $i$th gold center and the $j$th atom. The sum runs over all the atomic charges in the simulation cell to obtain the electrostatic potential.

\textbf{Ion Adsorption Number}. The adsorption number per gold atom used in Table~\ref{tbl:facet-cn} was calculated by counting the number of neighbored ions around each surface gold atom and then averaging the results over all gold atoms belonging to the same facet. The cutoff distances for counting neighbored Na$^+$ and Cl$^-$ were set to 3.5 and 4.5 \AA, respectively. These values were chosen as the location of the first peaks of the computed free energy profile (corresponding to the first minimum of the number density profile) of Na$^+$ and Cl$^-$ on the planar gold (111) surface.~\cite{Geada2018a} Several different combinations of cutoff distances were applied for testing (ranging from 3--4 \AA~for Na$^+$ and 4--5 \AA~for Cl$^-$), which all gave the same qualitative trends as in Table~\ref{tbl:facet-cn}. 

\section{ACKNOWLEDGMENTS}
This project has received funding from the European Research Council (ERC) under the European Union's Horizon 2020 research and innovation programme (Grant Agreement 646659-NANOREACTOR). M.K. acknowledges the financial support from the Slovenian Research Agency (contracts P1-0055 and J1-1701). The authors acknowledge support by the state of Baden-W{\"u}rttemberg through bwHPC and the German Research Foundation (DFG) through grant no INST 39/963-1 FUGG (bwForCluster NEMO).

\section{Supporting Information}
Effective surface property; spatial density distribution and polarization of AuNP in different solutions; radial number/charge density distribution and electrostatic potential.

\bibliography{refs}

\end{document}


\renewcommand \thefigure {S\arabic{figure}}
\setcounter{figure}{0}  
\renewcommand \thetable {S\arabic{table}}
\setcounter{table}{0}  

\begin{table}
	\centering
	\caption{Effective surface properties compared between gold models with and without polarizability. $Z_{\rm{eff}}^{\rm{MD}}$ represents the effective charge obtained from ionic cumulative charge at $R_{\rm{eff}}$ (cf. Figure S6--S8),  $Z_{\rm{eff}}^{\rm{DH}}$ and $\phi_{\rm{eff}}^{\rm{DH}}$ denote the effective charge and potential obtained from DH fitting. $^a$Values taken at larger radius ($\approx 3$ \AA~away from $R_{\rm{eff}}$) for Na$_3$HCF and Na$_4$HCF systems.} 
	\label{tbl:effective} %
	\scalebox{1}{%
		\begin{tabular}{ccccccccccc}
			\hline 
			\rule{0pt}{4ex}NPs&Salt&$Z_{\rm{eff}}^{\rm{MD}}$ [$e_0$] &$Z_{\rm{eff}}^{\rm{MD}}$ [$e_0$]  &$Z_{\rm{eff}}^{\rm{DH}}$ [$e_0$] &$Z_{\rm{eff}}^{\rm{DH}}$ [$e_0$] &$\phi_{\rm{eff}}^{\rm{DH}}$ [mV] &$\phi_{\rm{eff}}^{\rm{DH}}$ [mV]  \\ 
			\rule{0pt}{1ex} & & (Aue) & (Au) & (Aue) & (Au) & (Aue) & (Au) \\
			\hline 
			\rule{0pt}{3ex} AuNP1&NaCl      &$-$0.20 &$-$0.27 &$-$0.3  &$-$0.5  &$-$2  &$-$3 \\
			\rule{0pt}{3ex} AuNP2&NaCl      &$-$0.11 &$-$0.38 &$-$0.4  &$-$0.5  &$-$2  &$-$3 \\
			\rule{0pt}{3ex} AuNP3&NaCl      &$+$0.31 &$-$0.48 &$-$0.4  &$-$0.9  &$-$1  &$-$2 \\
			\rule{0pt}{4ex} AuNP1&NaNip     &$-$11.3 &$-$13.9 &$-$13.6 &$-$14.6 &$-$78 &$-$83 \\
			\rule{0pt}{3ex} AuNP2&NaNip     &$-$12.6 &$-$14.3 &$-$15.9 &$-$16.1 &$-$80 &$-$81 \\
			\rule{0pt}{3ex} AuNP3&NaNip     &$-$30.6 &$-$36.2 &$-$35.8 &$-$32.7 &$-$79 &$-$70 \\
			\rule{0pt}{4ex} AuNP2&NaBF$_4$  &$-$0.85 &$-$1.4  &$-$0.95 &$-$1.6  &$-$6  &$-$8 \\
			\rule{0pt}{3ex} AuNP2&NaPF$_6$  &$-$2.2  &$-$2.3  &$-$4.3  &$-$3.9  &$-$21 &$-$19 \\
			\rule{0pt}{3ex} AuNP2&Na$_3$HCF &$-$6.9$^a$&$-$3.5$^a$&$-$9.9 &$-$5.3 &$-$45 &$-$22 \\
			\rule{0pt}{3ex} AuNP2&Na$_4$HCF &$-$2.4$^a$&$-$0.6$^a$&$-$4.5 &$-$3.2 &$-$23 &$-$15 \\
			[1ex]\hline 
		\end{tabular}\quad }
\end{table}

\begin{figure}[ht]
	\centering
	\includegraphics[width=0.3\linewidth]{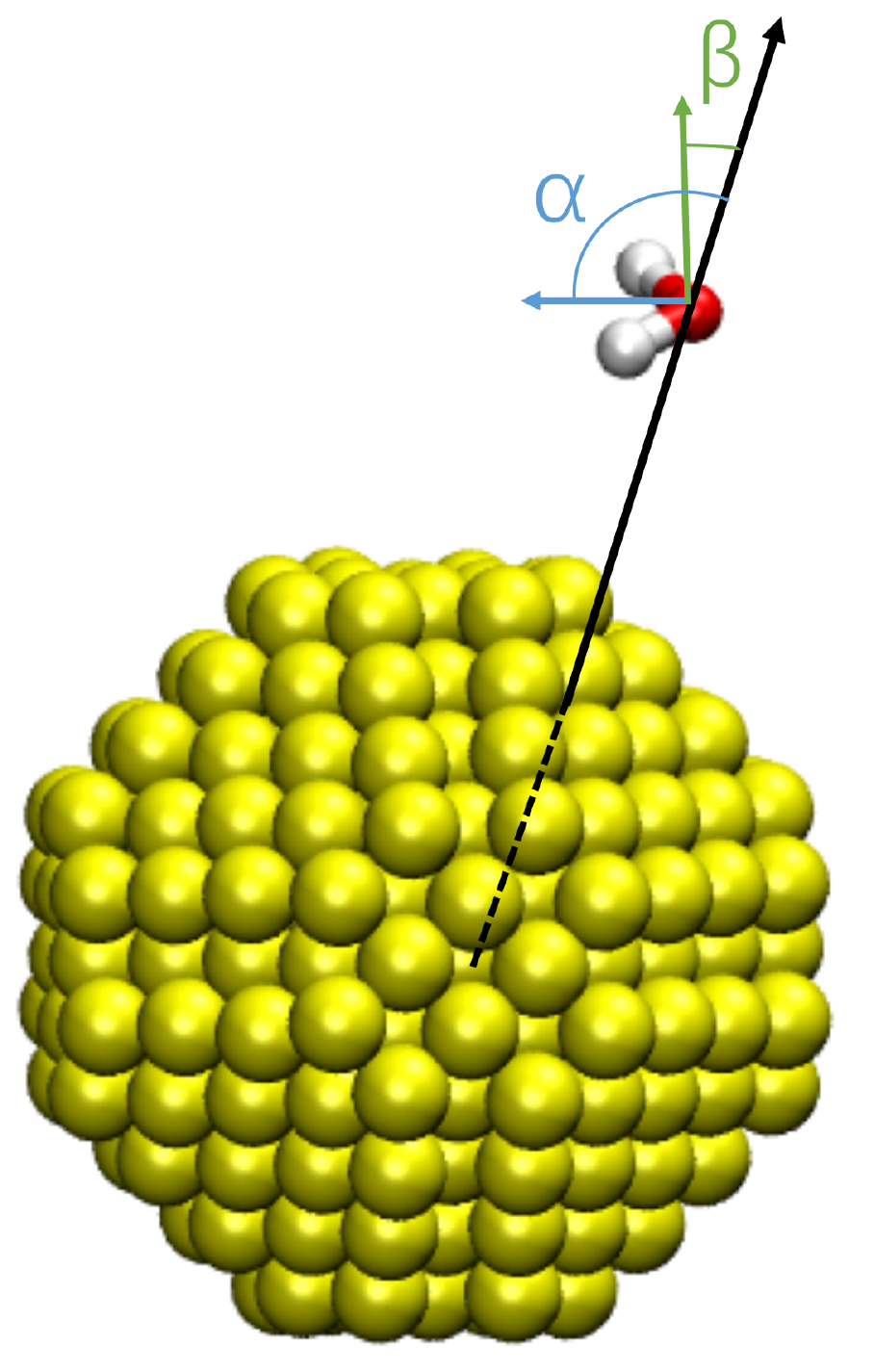}
	\caption{Schematic diagram for the definition of water orientation angles.}
	\label{figS:diagram}
\end{figure}

\begin{figure}[ht]
	\centering
	\includegraphics[width=1.\linewidth]{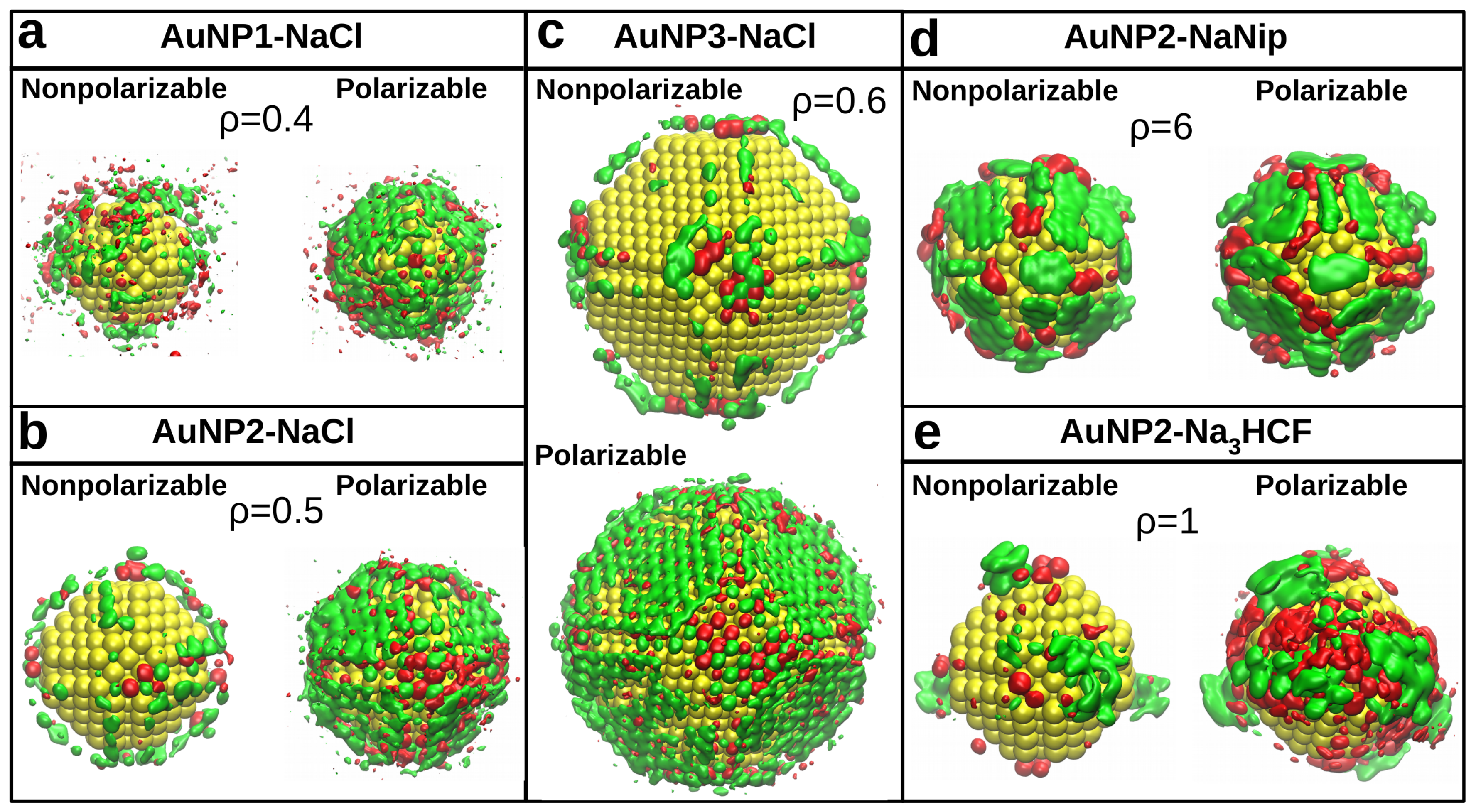}
	\caption{Spatial density distributions of ions around AuNPs for representative systems. Comparison between nonpolarizable and polarizable AuNPs is shown in each panel.}
	\label{figS:sdf}
\end{figure}

\begin{figure}[ht]
	\centering
	\includegraphics[width=1.\linewidth]{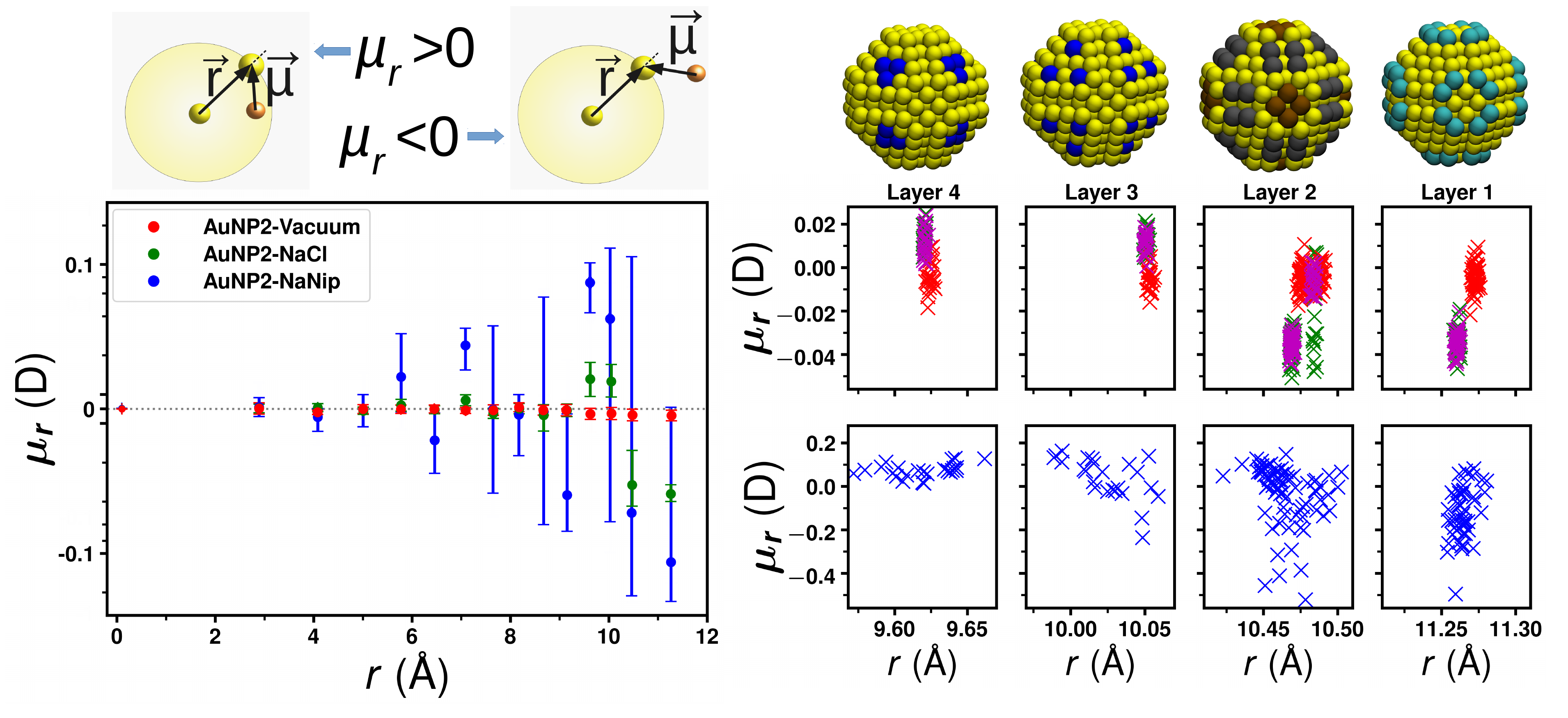}
	\put(-435,190){\textbf{\Huge a}}
	\put(-235,190){\textbf{\Huge b}}
	\caption{Polarization of AuNP2 in different media. (a) Averaged radial dipole moments as a function of $r$. The top panel shows illustrations of the positive and negative radial dipole moments. Discrete points represent the average values in each ``layer" and the bars show standard deviations in that layer. (b) Distributions of radial dipole moment at the surface separated by four layers according to distance $r$. The top panel highlights the selected atoms in each layer. Color codes in the lower panels are the same in panel (a), and the additional purple color shows the results for AuNP2 in pure water.
	For AuNP2 in vacuum, the overall average radial dipole moments are nearly zero because of the absence of external charges. For AuNP2 in the NaCl solution, differences are only obviously observed in the outer layers ($r >$ 9~\AA) compared with that in vacuum. Specifically, the last two layers show relatively larger negative values and the second last two layers have positive radial dipole moments. In contrast, drastic changes are observed for AuNP2-NaNip system, where much larger values (either negative or positive) are found in the outer layers along with large deviation in each layer, and even further impacted  inner layers. 
    Most significant differences among AuNP in different media are observed in the last four layers ($r >$ 9~\AA), which corresponds to gold atoms at the NP surfaces. Therefore, it is necessary to check the individual dipole moment distribution in these layers. We plot them separately and label them as Layer 1 to 4 as a gradually decreasing radial distance. Layer 1 corresponds to the vertex of the AuNP2 atoms, Layer 2 is a mixture of gold atoms on the (100) facets and the rest of edgy the (110) facets, Layer 3 and Layer 4 are gold atoms on the (111) facets. We can see that AuNP2-NaNip system shows the largest deviation of both the dipole moment and the gold atom position compared with other systems. The only difference between AuNP2-water (purple) and AuNP-NaCl (green) systems is observed in Layer 2 (more specifically, atoms on the (100) facets, which have slightly larger distance than those on the (110) facets), where part of gold atoms have more negative dipole moments in the latter system.}
	\label{figS:dipole-1}
\end{figure}

\begin{figure}[ht]
	\centering
	\includegraphics[width=1\linewidth]{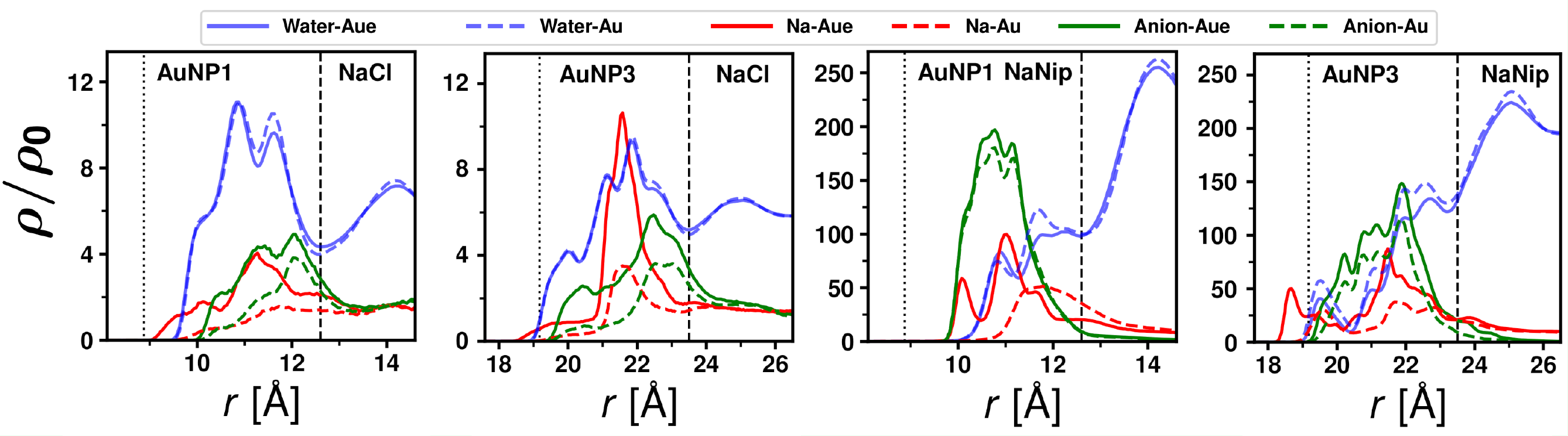}
	\caption{Radial number density distributions normalized by bulk density as a function of $r$ for AuNP-NaCl and AuNP-NaNip systems. Water profiles were scaled by a factor of 6 and 200, for AuNP-NaCl and AuNP-NaNip systems, respectively.}
	\label{figS:rdf1}
\end{figure}

\begin{figure}[ht]
	\centering
	\includegraphics[width=1\linewidth]{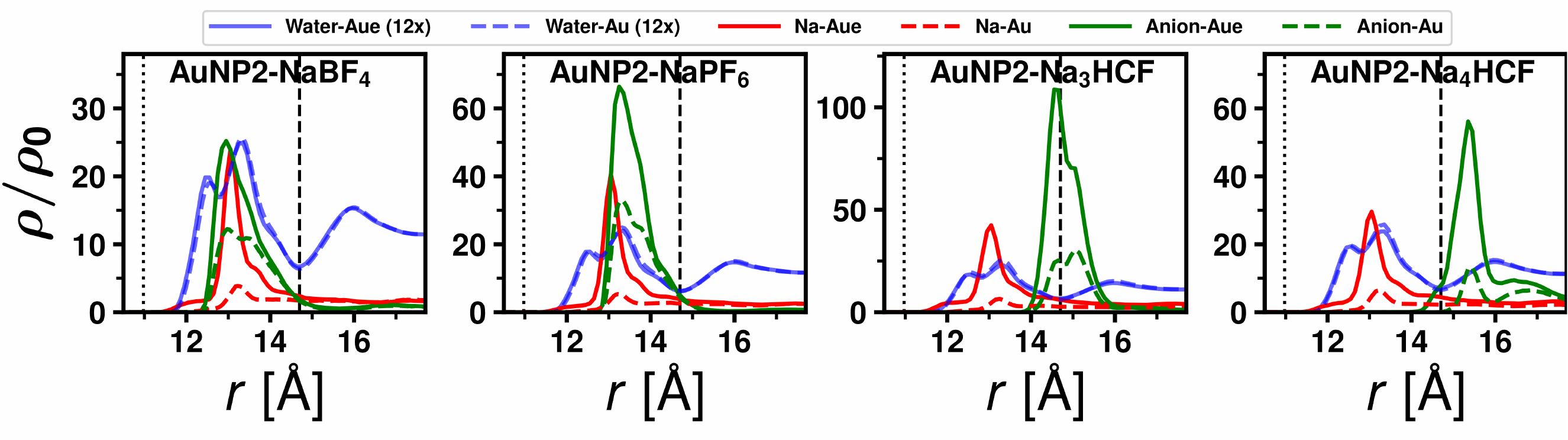}
	\caption{Radial number density distributions normalized by bulk density as a function of $r$ for AuNP2 with different salts. Note, water profiles were scaled by a factor of 12 for clarity.}
	\label{figS:rdf2}
\end{figure}

\begin{figure}[ht]
	\centering
	\includegraphics[width=1\linewidth]{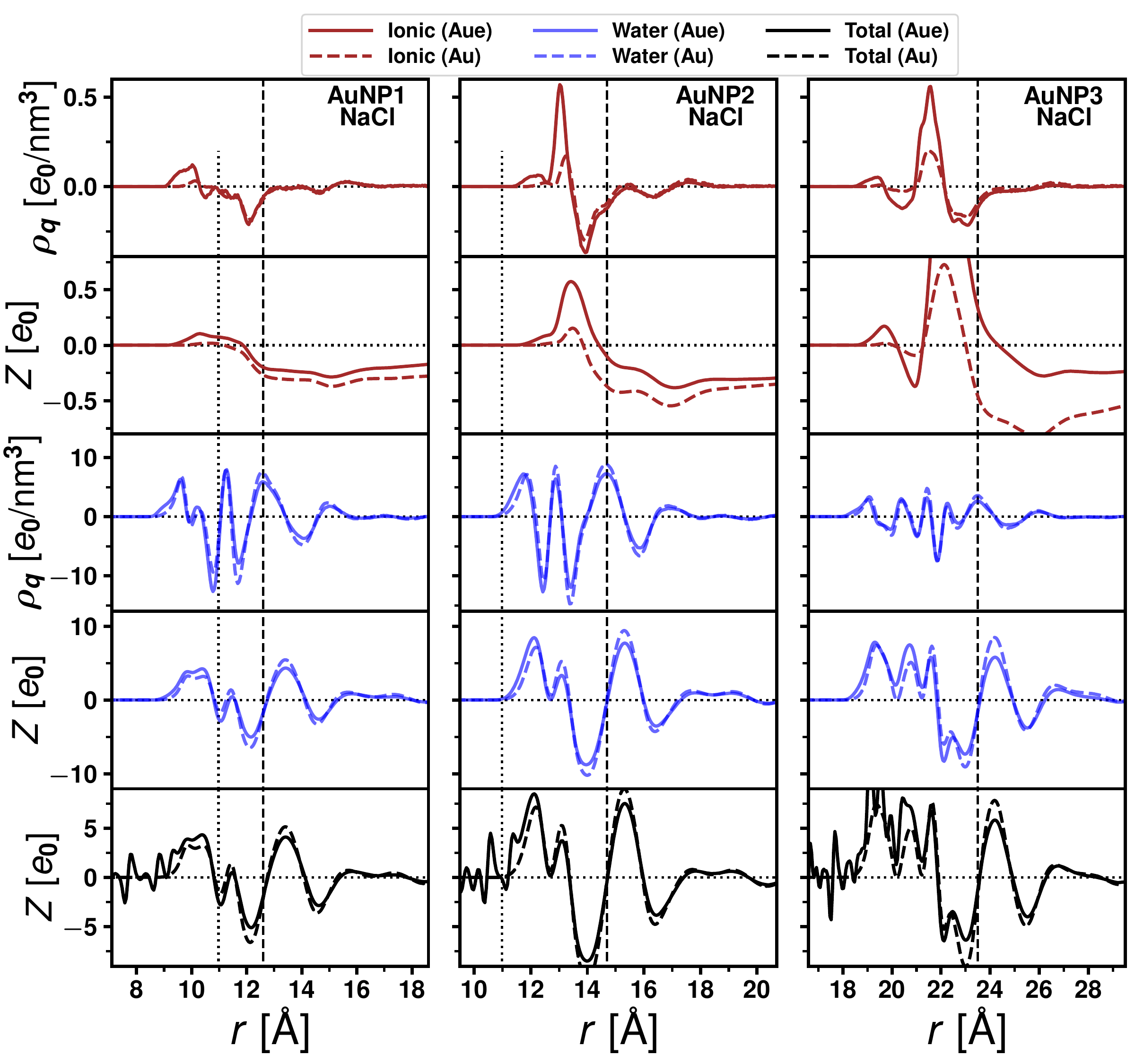}
	\caption{Charge density distributions ($\rho_q$) and accumulated charges ($Z$) as a function of $r$ for AuNPs in NaCl solutions. Decomposed properties into the ionic part (sum of cation and anion) and water contributions are shown in the first four rows and the total accumulated charge of the corresponding systems are shown in the lowest row.}
	\label{figS:charge-NaCl}
\end{figure}

\begin{figure}[ht]
	\centering
	\includegraphics[width=1\linewidth]{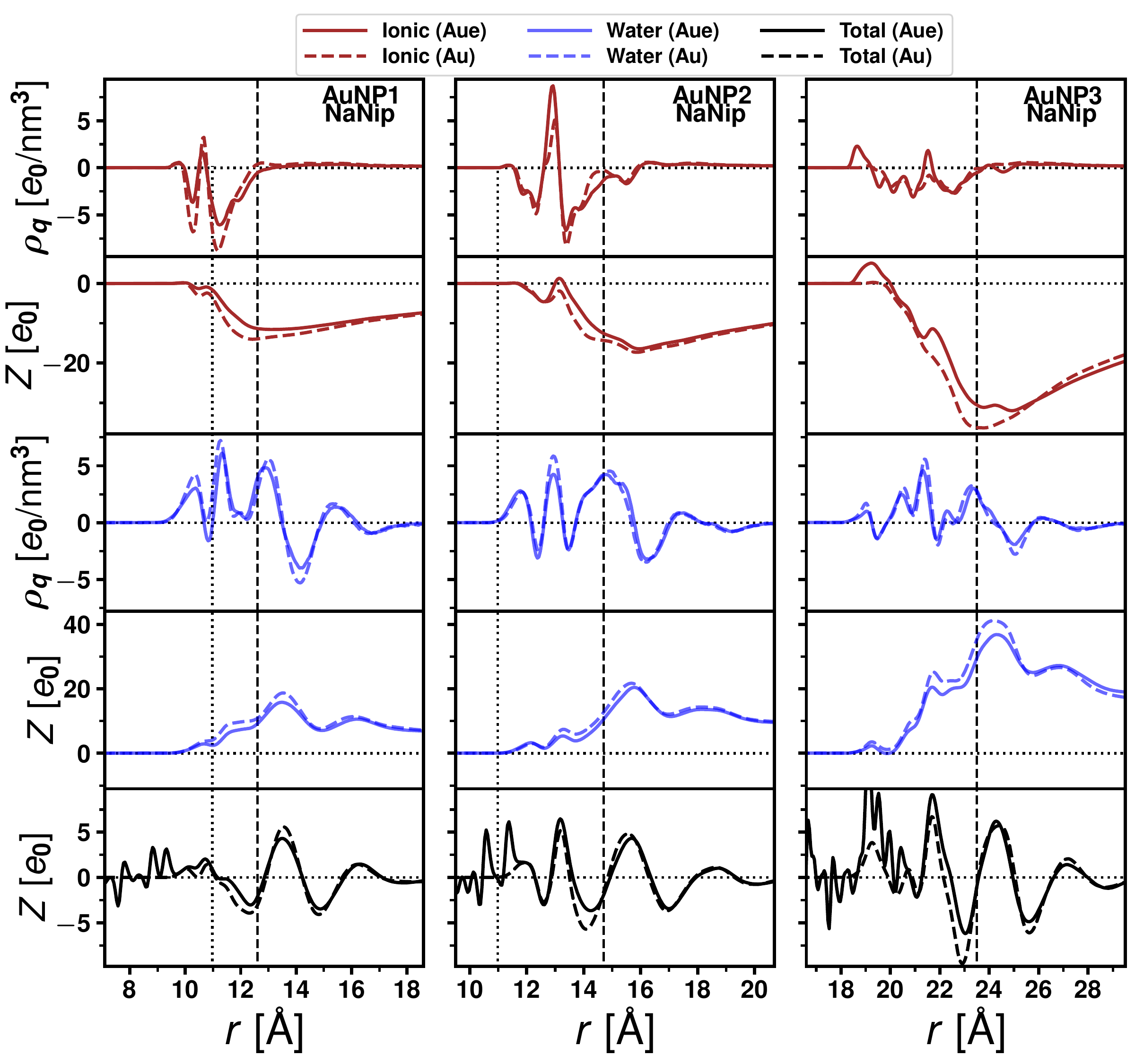}
	\caption{Charge density distributions and accumulated charges as a function of $r$ for AuNP-NaNip systems.}
	\label{figS:charge-NaNip}
\end{figure}

\begin{figure}[ht]
	\centering
	\includegraphics[width=1\linewidth]{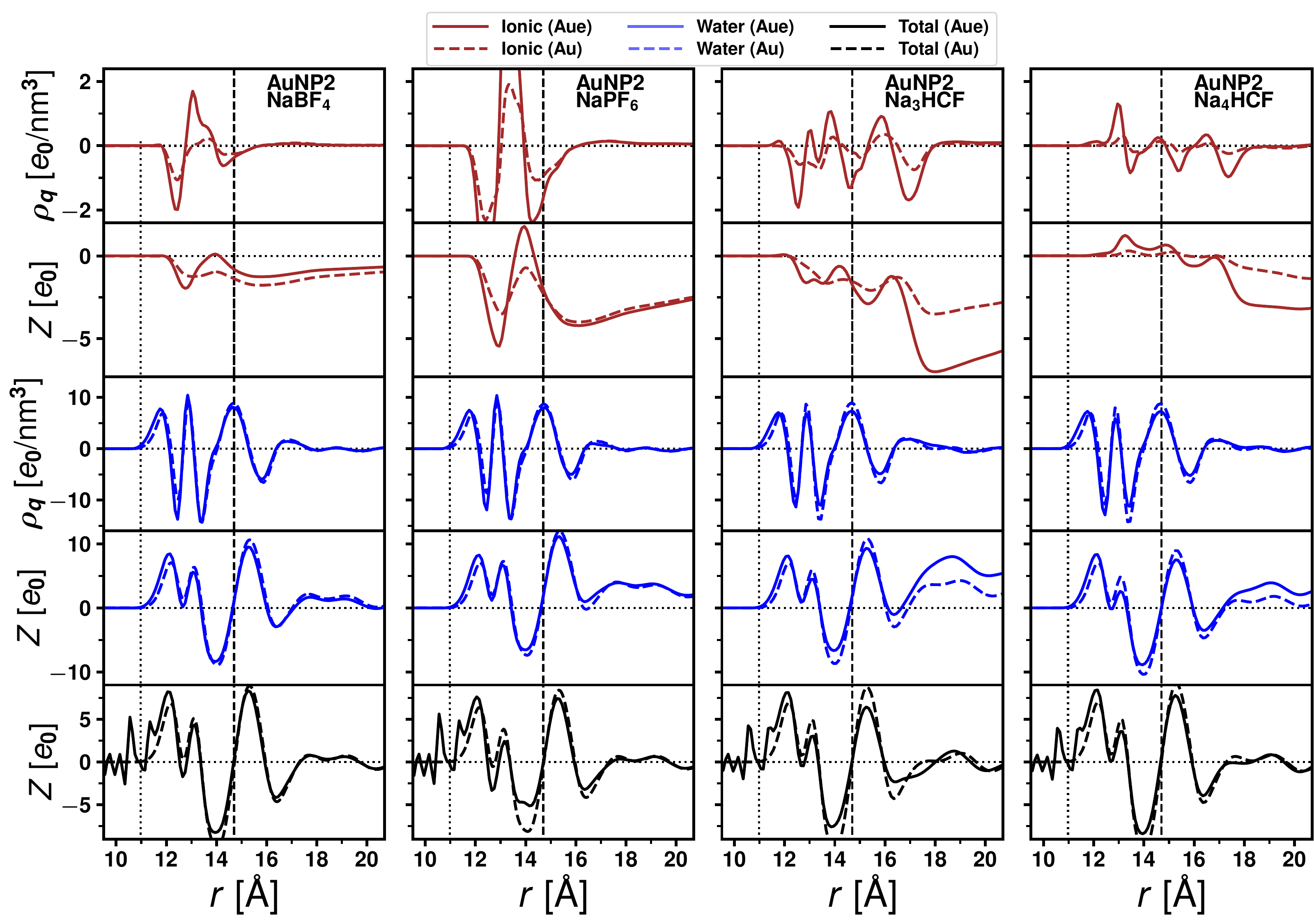}
	\caption{Charge density distributions and accumulated charges as a function of $r$ for AuNP2 with other salt systems.}
	\label{figS:charge-other}
\end{figure}

\begin{figure}[ht]
	\centering
	\includegraphics[width=1\linewidth]{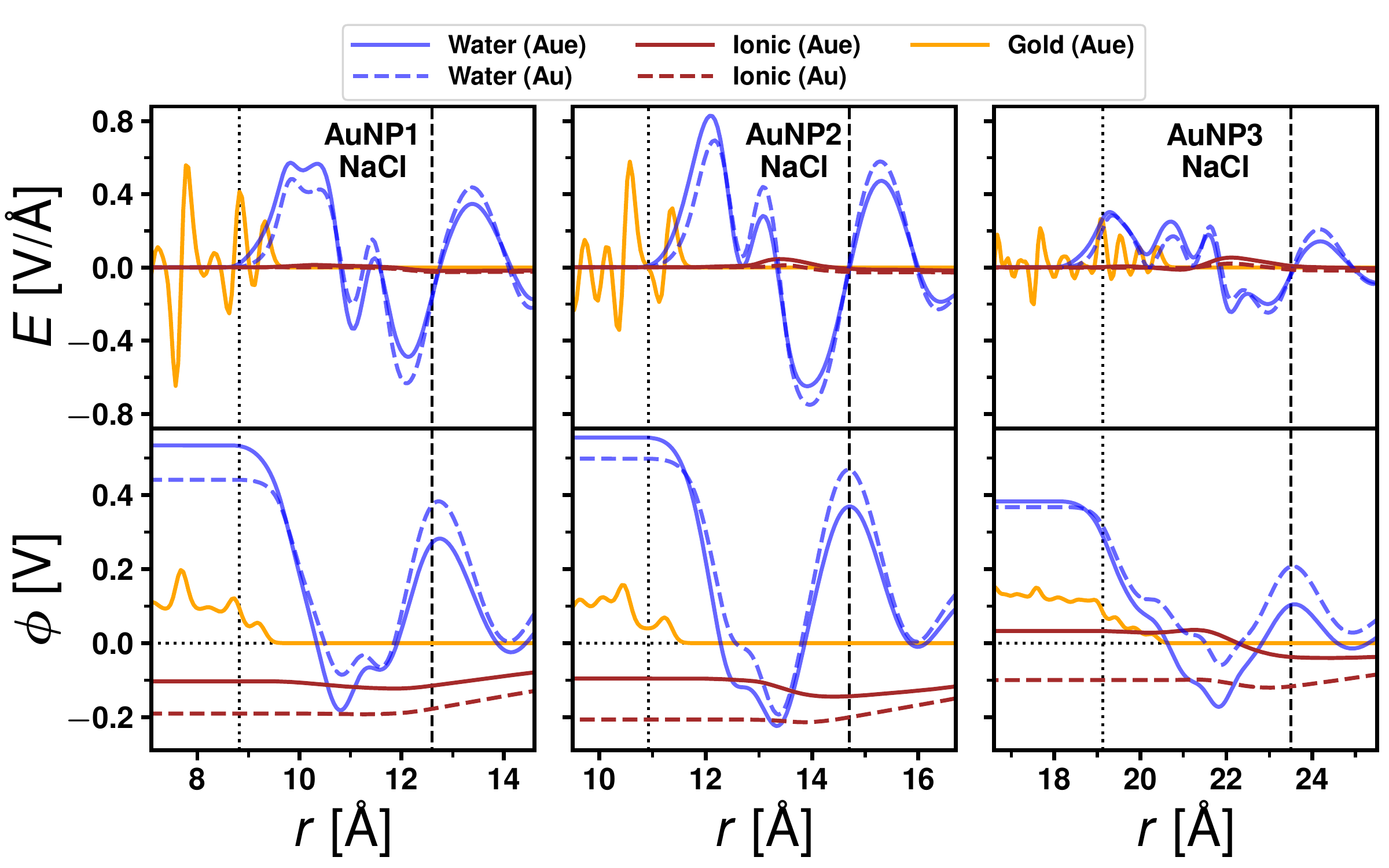}
	\caption{Decomposed electric field $E$ and potential $\phi$ as a function of $r$ for AuNP-NaCl systems. Comparisons between polarizable and nonpolarizable NPs were distinguished by solid and dashed lines.}
	\label{figS:electro-NaCl}
\end{figure}

\clearpage
\begin{figure}[ht]
	\centering
	\includegraphics[width=1\linewidth]{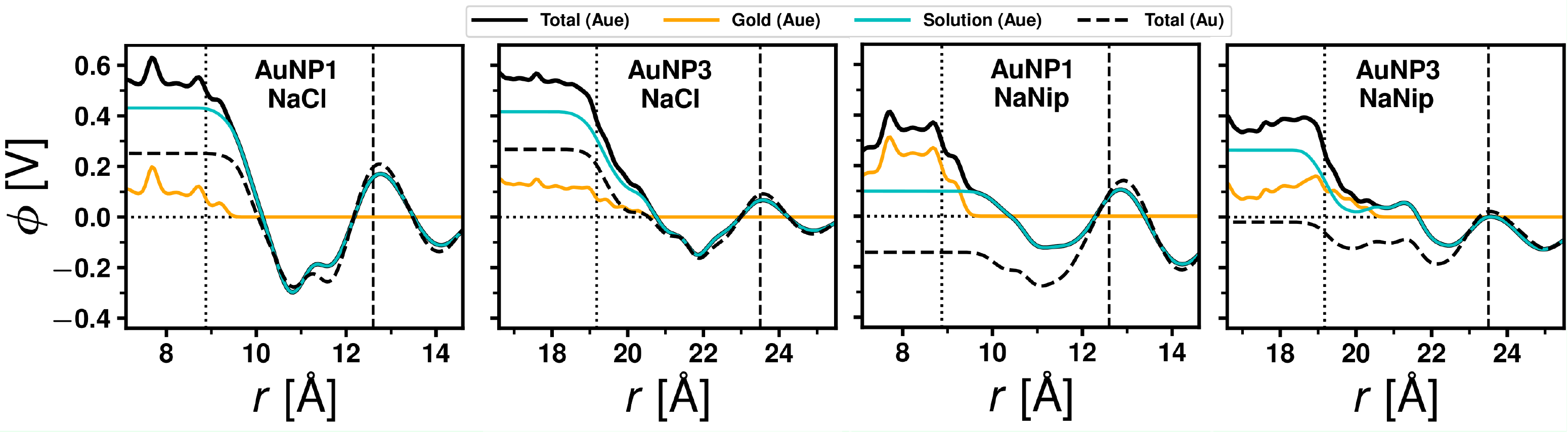}
	\caption{Total and decomposed electrostatic potential profiles as a function of $r$ for AuNP-NaCl and AuNP-NaNip systems. Total potential obtained from nonpolarizable gold model is shown by dashed curves for comparison.}
	\label{figS:pot1}
\end{figure}

\begin{figure}[ht]
	\centering
	\includegraphics[width=1\linewidth]{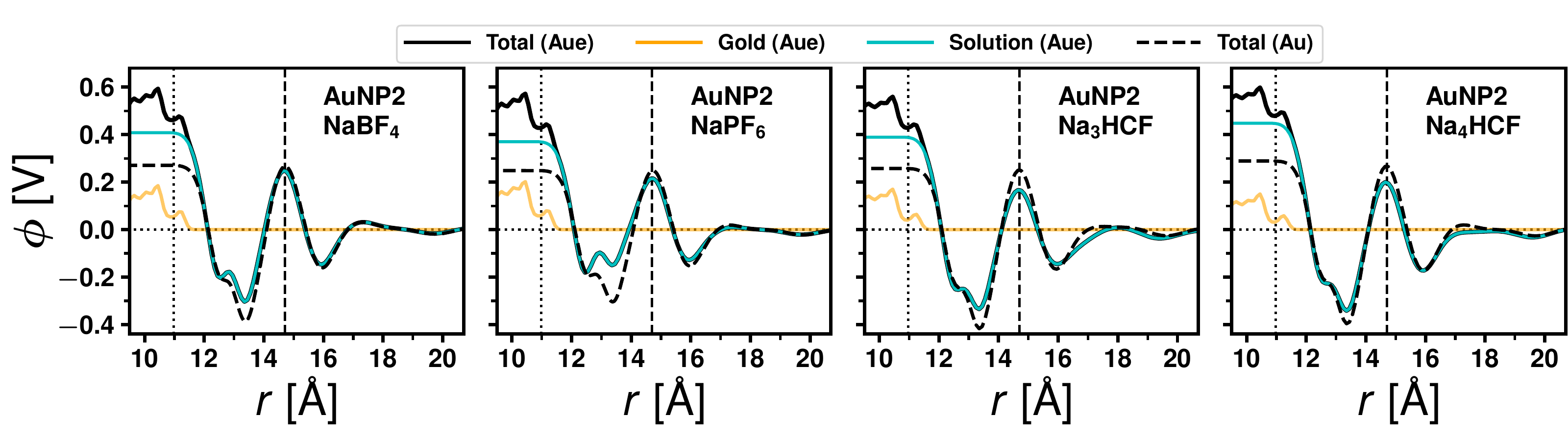}
	\caption{Total and decomposed electrostatic potential profiles as a function of $r$ for AuNP2 with different salt systems. Total potential obtained from nonpolarizable gold model is shown by dashed curves for comparison.}
	\label{figS:pot2}
\end{figure}